\documentclass[]{aa}
\usepackage{psfig}
\begin{document}

\title{The multimode pulsation of the $\delta$ Scuti star
V784~Cassiopeae\thanks{Based on spectroscopic data obtained at the David
Dunlap Observatory, University of Toronto and photometric observations
obtained at the Sierra Nevada Observatory, IAA-CSIC}}

\author{L.L. Kiss\inst{1,4,6} \and A. Derekas\inst{1,4} \and E.J. Alfaro\inst{2} \and
I.B. B\'\i r\'o\inst{3,4} \and B. Cs\'ak\inst{1,4} \and R. Garrido\inst{2}
\and K. Szatm\'ary\inst{1} \and J.R. Thomson\inst{5} }

\institute{Department of Experimental Physics and Astronomical Observatory,
University of Szeged, Szeged, D\'om t\'er 9., H-6720 Hungary \and
Instituto de Astrof\'\i sica de Andaluc\'\i a, CSIC, P.O. Box 3004,
E-18080, Spain \and
Baja Astronomical Observatory of B\'acs-Kiskun County, Baja, P.O. Box 766,
H-6501 Hungary \and
Guest Observer, Sierra Nevada Observatory, IAA-CSIC \and
David Dunlap Observatory, University of Toronto, Richmond Hill, Canada
\and 
Hungarian E\"otv\"os Fellowship, Instituto de Astrof\'\i sica de Andaluc\'\i a,
CSIC}

\titlerunning{The $\delta$ Scuti star V784~Cas}
\authorrunning{Kiss et al.}
\offprints{L.L. Kiss, \email{l.kiss@physx.u-szeged.hu}}
\date{}

\abstract{
We present an analysis of new Johnson and Str\"omgren photometric and medium-resolution
spectroscopic observations of the $\delta$ Scuti type variable star
V784~Cassiopeae. The data were obtained in three consecutive years between
1999 and 2001. The period analysis of the light curve 
resulted in the detection of four frequencies ranging from 9.15 d$^{-1}$ 
to 15.90 d$^{-1}$, while there is a suggestion for more, unresolved 
frequency components, too. The mean Str\"omgren indices and Hipparcos 
parallax were combined to calculate the following physical parameters:
$\langle T_{\rm eff} \rangle$=7100$\pm$100 K,
${\rm log}~g$=3.8$\pm$0.1, $M_{\rm bol}$=1\fm50$\pm$0\fm15.
The position of the star in the HR diagram was used to derive evolutionary
mass and age yielding to a consistent picture of an evolved $\delta$ Scuti
star with a mixture of radial plus non-radial modes.
}

\maketitle

\section{Introduction}

$\delta$ Scuti-type variable stars are pulsating variables located in the lower
part of the classical instability strip near or slightly above the
main-sequence. The characteristic time-scale of the light variation is in the
order of 0\fd1, while  the observed light curves are usually multiperiodic due
to the simultaneously excited radial and/or non-radial modes. Reliable mode
identification requires long-term and/or multi-site observing campaigns, which
are well-illustrated, e.g. by recent results from the Delta Scuti Network
(Breger et al. 1998, Breger 2000), STEPHI program (Alvarez et al. 1998) or the
Whole Earth Telescope (Handler et al. 1997). A handbook of reviews and
discussion of the astrophysical importance of these variables has been
published very recently (Breger \& Montgomery 2000). The most complete
catalogue of $\delta$ Scuti stars has been tailored and analysed by 
Rodr\'\i guez et al. (2000) and Rodr\'\i guez \& Breger (2001).

The light variation of the short period variable V784~Cas
(=HD~13122=BD+59$^\circ$422, P=0.1092 d, $\Delta$V=0\fm06, spectral type
F5II, ESA 1997) was discovered by the Hipparcos astrometric satellite. The
R00 catalogue (Rodr\'\i guez et al. 2000) includes this star listing the
parameters derived from the Hipparcos observations. The star lies about
1$^\circ$ NW of the open cluster Stock~2, but it is not associated with
this strongly reddened cluster located at 316 pc (Krzemi\'nski \& Serkowski
1967). The Hipparcos parallax (9.81$\pm$0.75 mas) supports the close
proximity of the star (102$^{+8}_{-7}$ pc). A few radial velocity
measurements can be found in the literature, they range from $-$6
km~s$^{-1}$ (De Medeiros \& Mayor 1999) to $+20$ km~s$^{-1}$ (Duflot et al.
1995). $UBVRI$ photometry was given by Fernie (1983), while the star was
included in the list of bright northern stars with interesting Str\"omgren
indices by Olsen (1980). V784~Cas was also studied in a sample of bright
giant stars by L\`ebre \& De Medeiros (1997), where no emission features,
neither time variations or asymmetries of the H$\alpha$ line profiles have
been detected (this star was observed two times separated by ten months).
The measured rotational velocity is 66 km~s$^{-1}$ (De Medeiros \& Mayor
1999). Neither of the studies mentioned above dealt with the time-dependent
phenomena, only the scatter of the velocity measurements (4 km~s$^{-1}$), as
listed in De Medeiros \& Mayor (1999), suggested the possible variability.
Most recently, Gray et al. (2001) included the star in their large sample of
late A-, F- and early G-type stars and determined its spectral type
(F0-F2III). They also noted that V784~Cas is a mild A$m$ star, the lines of
Sr II $\lambda$4077 and $\lambda$4216 are enhanced. There is no metallicity
determination in the literature.

We started a long-term observational project of obtaining follow-up
observations of bright, new variable stars discovered by the Hipparcos
satellite. We have so far identified a candidate second overtone field
RR~Lyrae variable (Kiss et al. 1999a), a new high-amplitude $\delta$ Scuti
star (Kiss et al. 1999b) and revealed the misclassification of a contact
binary (Cs\'ak et al. 2000). The main aim of this paper is to present an
analysis of new photometric and spectroscopic observations of V784~Cas. The
paper is organised as follows: the observations are described in Sect.\ 2,
Sect.\ 3 deals with the period analysis, while radial velocities are
discussed in Sect.\ 4. Finally, the physical parameters 
are presented in Sect.\ 5.

\section{Observations}

\subsection{Photometry}

Photoelectric Johnson photometry was carried out on 6 nights in
September-December, 1999 using the 0.4-m Cassegrain telescope
of Szeged Observatory (Hungary). The detector was a single-channel
SSP-5A photoelectric photometer. We made differential
photometry with respect to HD~14172 ($V$=6\fm96, $B-V$=+0\fm22,
$U-B$=+0.20, Krzemi\'nski \& Serkowski, 1967), which is the bright member
of the visual double star BDS~1193. The check star was HD~14173
($V$=7\fm20, $B-V$=+1\fm00, $U-B$=+0\fm66, Krzemi\'nski \& Serkowski, 1967),
the faint member of the system, located at $\sim$1$^\prime$ from the
comparison star. Since the diaphragm of the photometer is
30$^{\prime\prime}$, we could well measure the individual stars. We note,
that $U$-band observations were carried out only on the last two nights of
the observing run.

\begin{figure}
\begin{center}
\leavevmode
\psfig{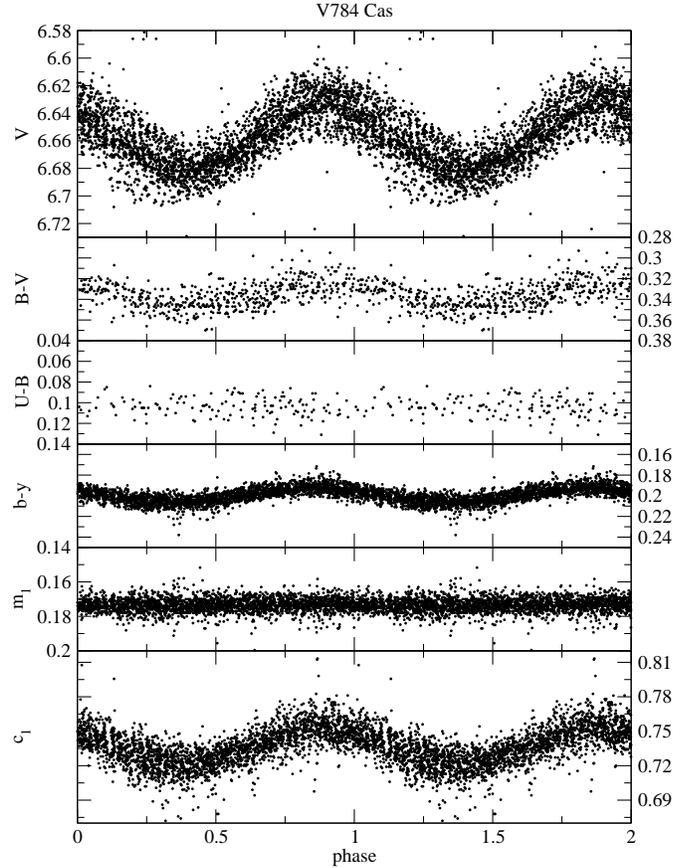}
\caption{The light and colour curves phased with the
Hipparcos ephemeris. Note the $\sim-$0.1 shift of the maximum
indicating the outdated light elements.}
\end{center}
\label{f1}
\end{figure}

\begin{table}
\caption{The journal of observations}
\begin{center}
\begin{tabular} {|lll|lll|}
\hline
J.D. & type & hours & J.D. & type & hours \\
\hline
2451433  & $BV$ &  3.4 &      2452153  & $uvby$ & 4.1\\
2451435  & $BV$ &  3.4 &      2452154  & $uvby$ & 6.3\\
2451452  & $BV$ &  5.2 &      2452155  & $uvby$ & 2.3\\
2451469  & $UBV$ & 3.3 &      2452156  & $uvby$ & 5.6\\
2451480  & spectr. & 4.9 &    2452157  & $uvby$ & 4.1\\
2451519  & $UBV$ & 4.4 &      2452158  & $uvby$ & 5.2\\
2451781  & $uvby$ & 4.7 &     2452159  & $uvby$ & 5.4\\
2451785  & $uvby$ & 1.0  &    2452160  & $uvby$ & 2.2\\
2451788  & $uvby$ & 4.0 &     2452161  & $uvby$ & 6.3\\
2451789  & $uvby$ & 1.0 &     2452162  & $uvby$ & 7.3\\
2452151  & $uvby$ & 4.2 &     2452191  & spectr. & 2.4\\
2452152  & $uvby$ & 5.4 &     2452200  & spectr. & 5.3\\
\hline
\end{tabular}
\end{center}
\end{table}

The Str\"omgren $uvby$ photometric observations were acquired on 4 nights in
August and September, 2000 and 12 nights in August and September, 2001 using
the 0.9-m telescope of the Sierra Nevada Observatory (Spain) equipped with a
four-channel spectrograph photometer. The differential photometric data were
obtained using the same comparison and check stars (HD~14172: $b-y$=0\fm154,
$m_1$=0\fm111, $c_1$=1\fm153; HD~14173: $b-y$=0\fm642, $m_1$=0\fm257,
$c_1$=0\fm455, Olsen 1993). The brightness and colour differences of the
comparison stars have been found to be constant to $\pm$0\fm01 (suggested
by the rms of the data). The
overall accuracy of the standard transformations is estimated to be about
$\pm$0\fm01 for the $V$ (both from the Johnson and Str\"omgren data), $B-V$ and
$b-y$ data, $\pm$0\fm015 for $U-B$ and $m_1$ and $\pm$0.02 for the $c_1$ data.
Because of the same comparison, the standardized $V$ light curves from the
Johnson and Str\"omgren measurements are well comparable (the
observed light range -- 6\fm60-6\fm70 -- was the same for both datasets). The
journal of observations is presented in Table\ 1.

We have obtained 3125 individual $V$ points, (512 from $UBV$ and 2613 from
$uvby$), 512 $B-V$ points, 132 $U-B$ points and 2613 Str\"omgren indices
\footnote{Individual data are available electronically at the CDS}.
The whole dataset has been phased with the Hipparcos ephemeris
(P=0\fd1092130, E=2448500.0700, ESA 1997) and the resulting phase
diagrams are plotted in Fig.\ 1. The observed behaviour of the
colour variations is typical in pulsating stars thus excluding the
possibility of other type of variation (e.g. eclipsing or ellipsoidal).
However, the light curve showed such cycle-to-cycle changes that
the assumption of the monoperiodic nature had to be rejected.

\subsection{Spectroscopy}

\begin{figure}
\begin{center}
\leavevmode
\psfig{figure=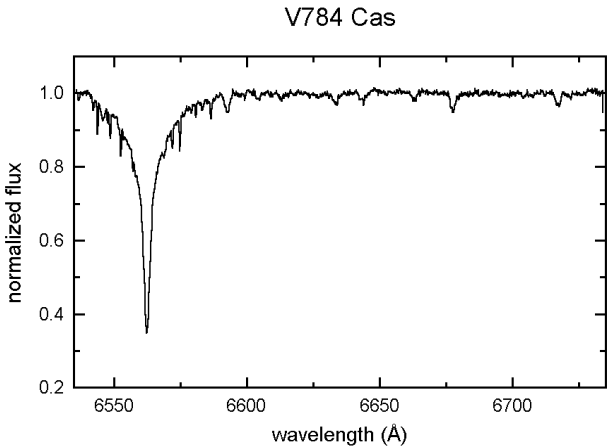,width=\linewidth}
\caption{Sample spectrum of V784~Cas}
\end{center}
\label{f2}
\end{figure}

The spectroscopic observations were carried out at the David Dunlap
Observatory with the Cassegrain spectrograph attached to the 74" telescope
on one night in October, 1999 and two nights in October, 2001. The detector
and the spectrograph setup were the same as used by Vink\'o et al. (1998).
The resolving power ($\lambda / \Delta \lambda$) was 11,000 and the
signal-to-noise ratio reached 70--100. The spectra in 1999 were centered on
6635 \AA, in 2001 on 6550 \AA\ and the wavelength span was 200 \AA\ in both
cases. The data were reduced with standard tasks in IRAF\footnote
{IRAF is distributed by
the National Optical Astronomy Observatories, which are operated by the
Association of Universities for Research in Astronomy, Inc., under  
cooperative agreement with the National Science Foundation.}, including bias
removal, flat-fielding, cosmic ray elimination, aperture extraction (with
the task $doslit$) and wavelength calibration. For the latter, two FeAr
spectral lamp exposures were used, which were obtained before and after
every five stellar exposures. Because of the short period of V784~Cas the
observing sequence of FeAr-var-var-var-var-var-FeAr was chosen. Careful
linear interpolation between the two comparison spectra was applied in order
to take into account the sub-pixel shifts of the five stellar spectra caused
by the tracking of the telescope. We chose an exposure time of 6 minutes,
which corresponds to $\sim$0.04 phase of the dominant period. The spectra
were normalized to the continuum by fitting a cubic spline, omitting the
region of H$\alpha$.

Besides the telluric features on the blue side of the
H$\alpha$ line, we could detect a few weak and broad metallic
lines. The H$\alpha$ profile remained symmetric during the observations
excluding the presence of high-amplitude non-radial oscillations. A sample
spectrum taken in 1999 is shown in Fig.\ 2. We have collected 98 individual
spectra for V784~Cas.

\section{Period analysis}

\begin{table}
\caption{The result of the period analysis. $f_4$ and $f_6$ seems 
to be caused by an unresolved frequency pair, 
thus discarded from further analysis.}
\begin{center}
\begin{tabular} {lrrr}
\hline
No. & freq. & ampl. & S/N\\
    & (d$^{-1}$) & (mmag) & \\
\hline
$f_1$ & 9.1565  & 24.3    &  27.6 \\
$f_2$ & 9.4649  & 8.4     &  9.5 \\
$f_3$ & 15.4036 & 5.5     &  8.3 \\
($f_4$) & 9.8800  & 4.9     &  5.6  \\
$f_5$ & 15.9013 & 3.5     &  5.3  \\
($f_6$) & 9.7777  & 3.3     &  4.1  \\
\hline
\end{tabular}
\end{center}
\end{table}

The period analysis was performed by means of standard Fourier-analysis
with subsequent prewhitening steps. For this we have used Period98
of Sperl (1998) which also includes multifrequency least squares
fitting of the parameters.

\begin{figure*}
\begin{center}
\leavevmode
\psfig{figure=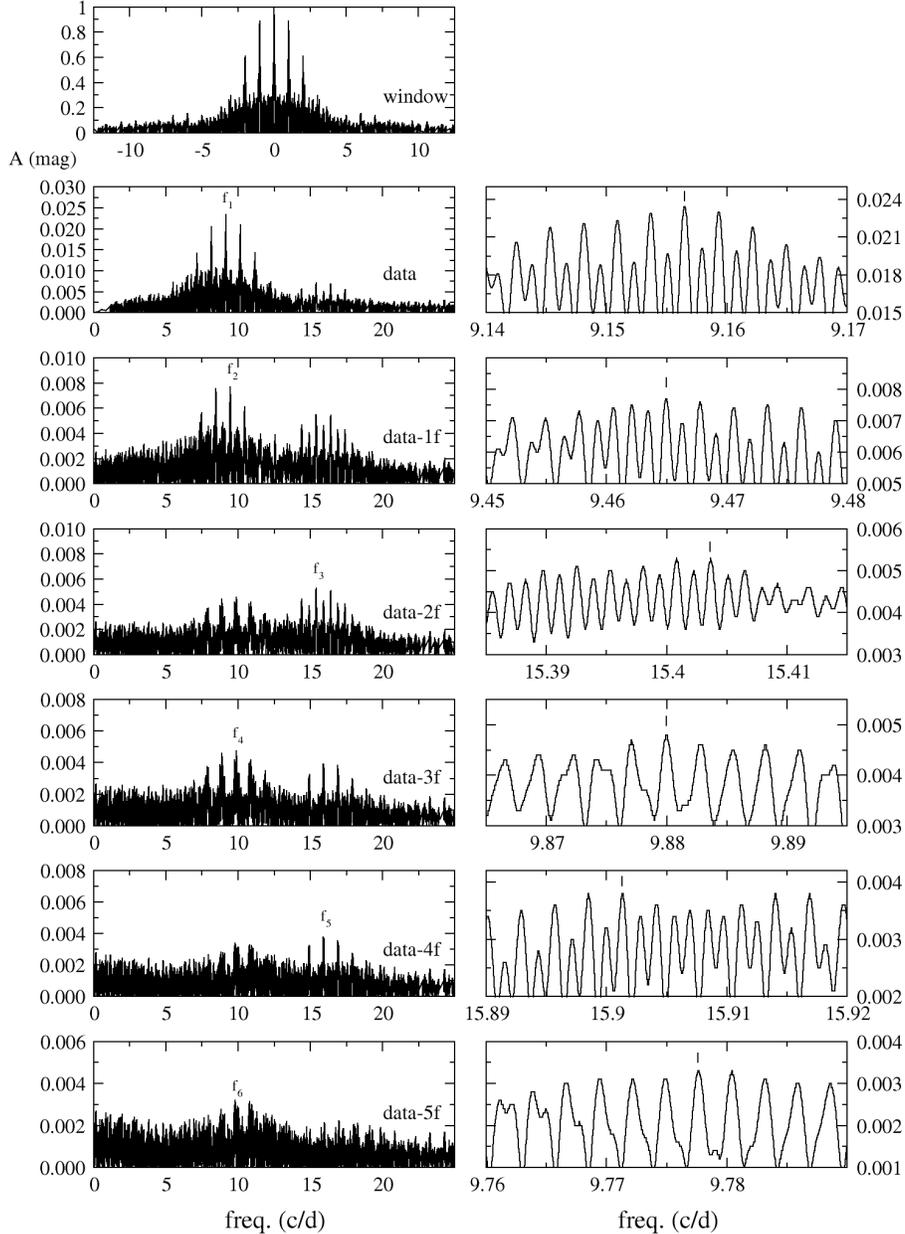,width=12cm}
\caption{Amplitude spectra of the whole $V$ light curves ({\it left column})
and close-up views of the main peaks ({\it right column}). Six
frequencies can be identified with S/N larger than 4.}
\end{center}
\label{f3}
\end{figure*}

First we decided to analyse the merged transformed data.
As has been mentioned above, the use of the same comparison star suggested 
a common analysis of the standard $V$-band data. The variable has
very similar $B-V$ and $b-y$ indices as the comparison, thus the
colour-dependent terms in the standard transformation equations are
only small corrections. Either the mean brightness (6\fm65) or the
extrema of the light curve are considered, the standard $V$ data
originated from Johnson-$BV$ and Str\"omgren-$by$ measurements agree
very well (the difference does not exceed 0\fm005). However, some spurious
low-frequency components yielded by the period analysis enforced us to 
reject this homogeneity assumption. During the prewhitening steps two
low-frequency components (at 0.0007 d$^{-1}$ and 0.5
d$^{-1}$) appeared suggesting: 1. a possible shift by a few millimagnitudes 
of the mean brightness from Johnson and Str\"omgren photometry and 2. possible slow
and low-amplitude variation of the comparison star mimicking changes of
the mean brightness on a daily basis. A close look at the comparison 
{\it minus} check magnitudes revealed indeed some slight changes of the 
daily averages with no systematic short-term variations. Therefore, 
we have adjusted the individual light curves (21 together) by 
subtracting nightly mean values (the differences are of order of 
a few millimagnitudes). Although 
the low-frequency components have been removed,
we have to admit that the mean values on some nights are fairly uncertain
because of shorter time-spans than mean periods. That is the reason why
the most critical subset obtained on JD 2451785 was excluded from
the frequency analysis.

The calculated amplitude spectra are presented in Fig.\ 3, where we show the
individual frequency spectra after consecutive prewhitenings.  
In order to illustrate the difficulty of the whole analysis we also 
present close-up views of the main peaks. Besides the strong 1/day alias
structure the 1/year is also strong. A certain amount of 
ambiguity cannot be excluded due to the complex spectral window
and finite spectral resolution.

The primary peak at $f_1=9.15650$ d$^{-1}$ is in very
good agreement with the Hipparcos result (9.15642 d$^{-1}$). 
In every step of the prewhitening 
procedure we allowed all of the parameters to vary to get the 
``best'' Fourier-fit of the light curve. 
In order to check the reality of the components, we have
determined the signal to noise ratios (S/N) following the suggestions of
Breger et al. (1993). The calculated S/N for the sixth components 
is 4.1 so it satisfies the proposed criterion of Breger et al. (S/N(real)$>$4)
for accepting ambiguous peaks in the frequency spectrum. 
Therefore, the final set consists of six frequencies
ranging from 9.15 c/d to 15.9 c/d and it is summarized in Table\ 2.

The period analysis has been checked by making use of the Hipparcos
Epoch Photometry data consisting of 117 points.
They were analysed separately and the calculated frequency spectrum (middle 
panel in Fig.\ 4) yields essentially to the same dominant period, that has
been used to phase the data (top panel in Fig.\ 4). After prewhitening with
this frequency, no further periodicity can be inferred from these data
(bottom panel in Fig.\ 4). A further consistency check is a comparison of 
Hipparcos and our dominant frequencies. It is shown in Fig.\ 5, where a
close-up to the main peaks is presented. As Hipparcos has good spectral
window (in sense of having only weak and asymmetric sidelobes), the
very good agreement of the main peaks supports our frequency analysis (see 
Jerzykiewicz \& Pamyatnykh 2000 for a recent discussion on the use of
Hipparcos photometric data).

Since the bulk of the data was obtained in 2001, a separate analysis of the
2001 data was carried out to yield insights into the reality of the
six-frequency solution. Furthermore, the noise in the Str\"omgren $v$
data is significantly reduced compared to transformed Johnson $V$ magnitudes
(i.e. the full amplitude is larger and the photometric accuracy is
better through the $v$ filter). The prewhitening and simultaneous nonlinear
parameter fitting resulted in largely similar frequencies at several 1/year
aliases of the finally adopted set (for instance, even the dominant 
component at 9.15 d$^{-1}$ is approximately a 1/year alias of the true
frequency). That is why we favoured frequencies resulted from the whole 
two-years long data.

\begin{figure}
\begin{center}
\leavevmode
\psfig{figure=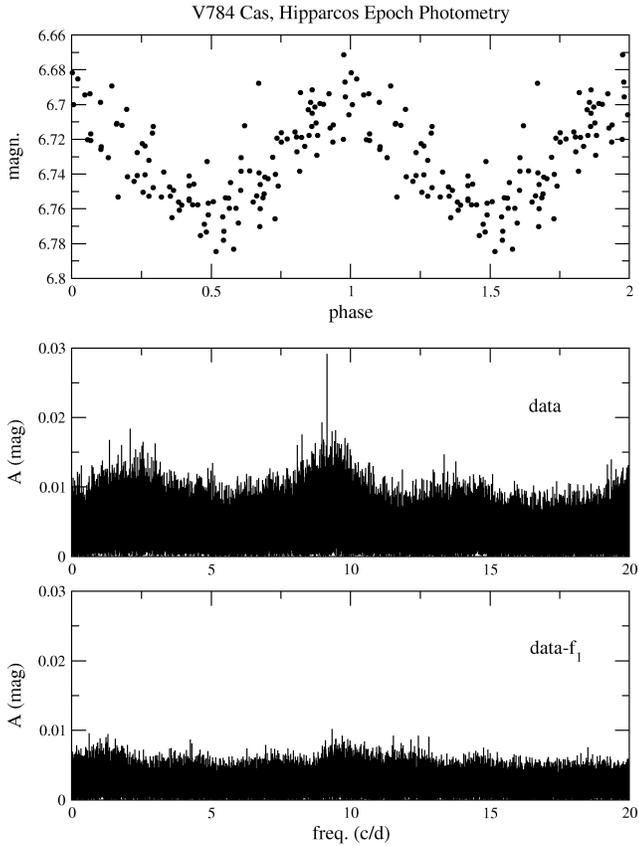,width=\linewidth}
\caption{The phase diagram of the Hipparcos data taken from the 
Hipparcos Epoch Photometry database calculated with a period of 
0.1092130 d (top panel) and the frequency spectrum (middle panel). 
No secondary period can be inferred from these data (bottom panel).}
\end{center}
\label{f4}
\end{figure}

\begin{figure}
\begin{center}
\leavevmode
\psfig{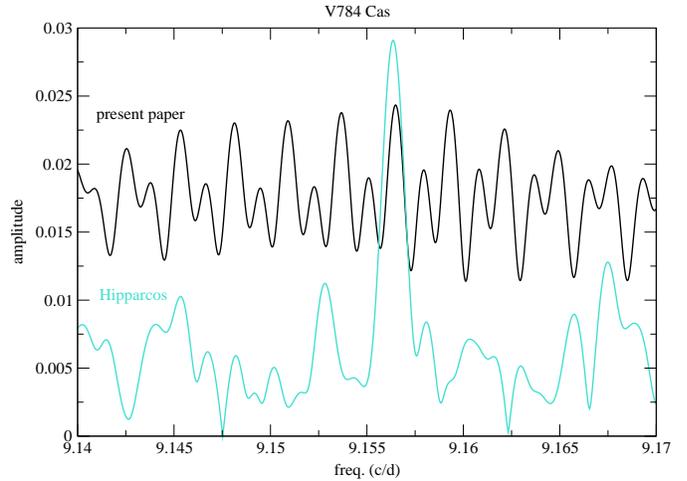}
\caption{A close-up to the main peaks in the frequency spectra calculated
from our data (1999-2001) and Hipparcos Epoch Photometry.}
\end{center}
\label{f5}
\end{figure}

Taking the six frequencies in Table\ 2, strong interaction between $f_4$ and
$f_6$ is obvious. A visual inspection of the corresponding close-up views
in Fig.\ 3 reveals that the peaks are broader than expected from the 
length of the data (the horizontal axes in the right column have the same
frequency scales). It is suggested that some very close components may 
exist which are beyond our spectral resolution (approximately $1.5/\Delta T$,
Loumos \& Deeming 1978). For our data $\Delta T$=729 d, which corresponds to
a $\Delta f\approx0.002$ d$^{-1}$. Most recently, Breger \& Bischof (2002)
studied close frequency pairs ($\Delta f<0.06$ d$^{-1}$) in $\delta$ Scuti
stars. A detailed discussion of the behaviour of BI~CMi led these authors
to conclude that very close frequency pairs do indeed exist and their presence
should be taken into account when planning photometric observations with the
best available spectral resolution. We conclude that similar
close frequencies might also exist in V784~Cas making difficult to interpret 
the presently available data. As a result, we kept only four frequencies for 
further analysis. The observed individual $V$ light curves are compared with
the four-component harmonic fit in Fig.\ 6. We turn back to the adopted 
frequencies in Sect.\ 5.

\begin{figure*}
\begin{center}
\leavevmode
\psfig{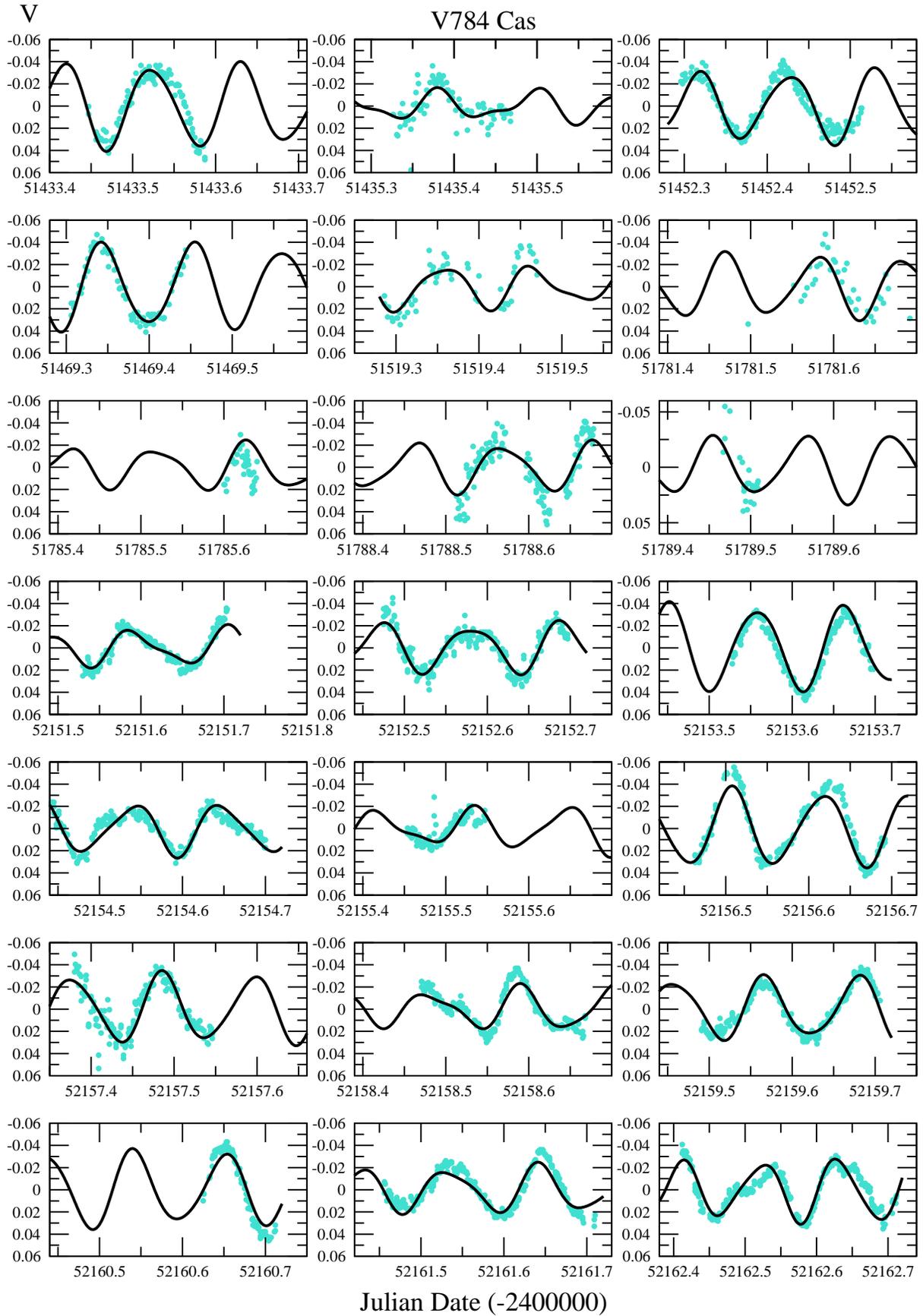}
\caption{The observed individual $V$ light curves (gray dots) with the 
four-component harmonic fit.}
\end{center}
\label{f6}
\end{figure*}

\section{Radial velocity variations}

\begin{table*}
\caption{Heliocentric radial velocities (km~s$^{-1}$) of V784~Cas 
(MJD=Hel. JD$-$2400000)}
\begin{center}
\begin{tabular} {lrlrlrlrlr}
\hline
MJD & $v_{\rm rad}$ & MJD  & $v_{\rm rad}$ & MJD & $v_{\rm rad}$ & MJD & $v_{\rm rad}$ & MJD & $v_{\rm rad}$ \\
\hline
51480.6788 &   $-$6.69 & 51480.8135 &   $-$6.15 & 52191.7601 &   $-$3.38 & 52200.7600 &   $-$7.05 & 52200.8462 &   $-$8.45 \\
51480.6878 &   $-$7.92 & 51480.8192 &   $-$5.40 & 52191.7730 &   $-$7.78 & 52200.7648 &   $-$8.22 & 52200.8499 &   $-$8.86 \\
51480.6922 &   $-$7.35 & 51480.8236 &   $-$7.45 & 52191.8006 &  $-$10.85 & 52200.7685 &   $-$6.64 & 52200.8536 &  $-$11.27 \\
51480.6966 &  $-$11.31 & 51480.8280 &   $-$8.64 & 52191.8063 &   $-$9.38 & 52200.7722 &   $-$5.25 & 52200.8573 &  $-$10.41 \\
51480.7010 &  $-$11.88 & 51480.8324 &   $-$6.56 & 52191.8124 &   $-$5.72 & 52200.7784 &   $-$4.96 & 52200.8611 &   $-$9.06 \\
51480.7074 &  $-$10.93 & 51480.8368 &   $-$6.05 & 52191.8188 &   $-$6.02 & 52200.7822 &   $-$3.61 & 52200.8648 &   $-$7.36 \\
51480.7118 &  $-$10.06 & 51480.8433 &   $-$7.49 & 52191.8261 &   $-$3.51 & 52200.7864 &   $-$2.90 & 52200.8800 &   $-$2.52 \\
51480.7162 &   $-$8.20 & 51480.8477 &   $-$4.56 & 52191.8326 &   $-$1.84 & 52200.7903 &   $-$2.78 & 52200.8838 &    0.35 \\
51480.7206 &   $-$8.70 & 51480.8521 &   $-$4.20 & 52191.8395 &   $-$1.95 & 52200.7960 &   $-$1.31 & 52200.8875 &   $-$0.16 \\
51480.7250 &   $-$7.86 & 51480.8565 &   $-$1.88 & 52200.7150 &   $-$0.58 & 52200.7998 &   $-$2.79 & 52200.8912 &    1.69 \\
51480.7313 &   $-$6.34 & 51480.8609 &   $-$1.79 & 52200.7187 &   $-$2.37 & 52200.8035 &   $-$1.24 & 52200.8949 &    1.65 \\
51480.7357 &   $-$6.09 & 51480.8667 &   $-$2.64 & 52200.7225 &    0.60   & 52200.8081 &   $-$1.39 & 52200.8987 &     2.71 \\
51480.7401 &   $-$5.42 & 51480.8711 &   $-$1.91 & 52200.7262 &   $-$4.11 & 52200.8118 &   $-$1.17 & 52200.9040 &    1.58  \\
51480.7445 &   $-$5.96 & 51480.8755 &   $-$2.30 & 52200.7302 &   $-$4.89 & 52200.8155 &    0.54   & 52200.9077 &   $-$0.69 \\
51480.7489 &   $-$5.53 & 51480.8799 &   $-$2.82 & 52200.7339 &   $-$7.64 & 52200.8217 &    1.57   & 52200.9114 &    0.56 \\
51480.7730 &   $-$7.70 & 51480.8843 &   $-$6.00 & 52200.7389 &   $-$6.76 & 52200.8254 &   $-$1.26 & 52200.9152 &   $-$0.41 \\
51480.7949 &   $-$6.51 & 52191.7410 &   $-$3.66 & 52200.7413 &   $-$8.93 & 52200.8291 &    0.19   & 52200.9189 &   $-$0.41 \\
51480.8002 &   $-$5.07 & 52191.7463 &   $-$3.09 & 52200.7465 &   $-$7.10 & 52200.8329 &   $-$1.22 & 52200.9226 &   $-$3.73 \\
51480.8046 &   $-$5.14 & 52191.7510 &   $-$3.26 & 52200.7526 &   $-$8.66 & 52200.8366 &   $-$3.93 &            &            \\
51480.8091 &   $-$5.65 & 52191.7556 &   $-$3.20 & 52200.7563 &   $-$6.86 & 52200.8403 &   $-$5.69 &            &         \\
\hline      
\end{tabular}
\end{center}
\end{table*}

Radial velocity variation of V784~Cas was determined by
measuring Doppler-shifts of the H$\alpha$ line. This is not an
ideal choice because the line forming region of the H$\alpha$
line extends to a much wider region than the photosphere, e.g.
it may have strong chromospheric component in the line core
(L\`ebre \& De Medeiros 1997 and references therein). However, the
observed spectral region does not contain other strong lines, the
detected metallic lines are too weak for radial velocity determination.
And, as it will be shown below, they are substantially asymmetric
suggesting the presence of non-radial oscillation.

Since the H$\alpha$ line is symmetric due to its saturation,
the radial velocities were determined
by fitting a parabola to the lowest points of the line profile.
The barycentric corrections were calculated with the IRAF task
{\it rvcorr}. The observed velocities are presented in Table\ 3.
Their estimated accuracy is about $\pm$1 km~s$^{-1}$, which is
based on our earlier experiences when using the same equipment
for studying other bright variables observed at similar
S/N ratios (Kiss et al. 1999ab). As an independent check,
we have also determined line bisector velocities
at various levels (see Kiss \& Vink\'o 2000 for an
application of this technique in Cepheid variables).
The mean difference of the resulting data is $\approx$0.5 km~s$^{-1}$
with a standard deviation of 1.0 km~s$^{-1}$ (even the most deviant
points did not differ more than 2 km~s$^{-1}$ from the line-core
velocities).

\begin{figure}
\begin{center}
\leavevmode
\psfig{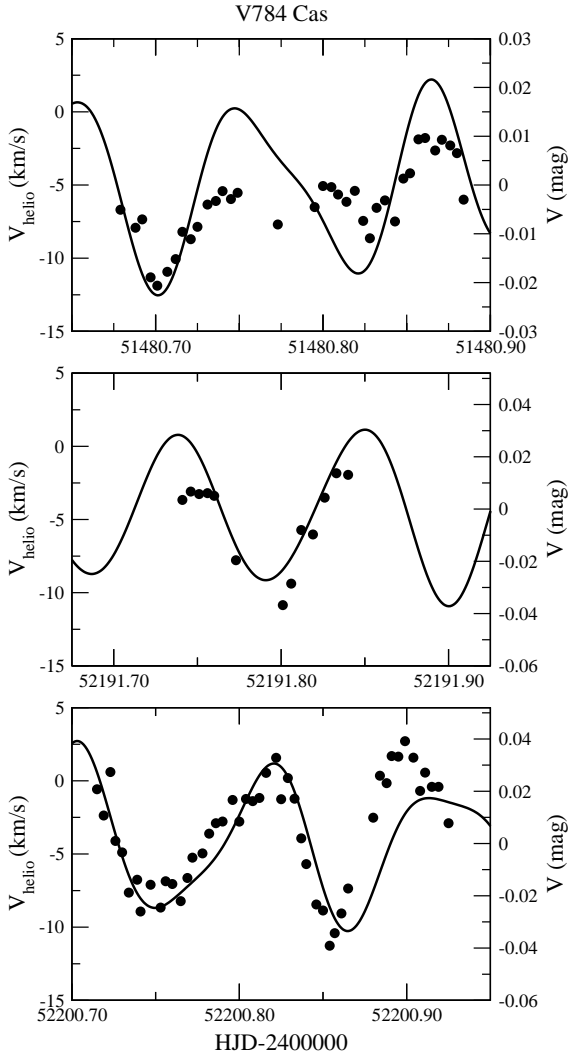}
\caption{Comparison of the radial velocity curves (dots) with the
 extrapolated light curve fits (solid line). Note the reversed scale of the
magnitude axes.}
\end{center}
\label{f7}
\end{figure}

The light and radial velocity variations have been compared using 
the light curve fit consisting of four frequencies. The
comparison is shown in Fig.\ 7. It is intended to illustrate 
the overall characteristics of the correlation between the
light and radial velocity variations and the 
ability of the four-frequency fit to predict light variation both 
as interpolation (JD 2451480) and extrapolation (JD 2452191 and 2452200).
We have estimated the value of $2K/\Delta V$ by taking the 
full amplitudes of the radial velocity curves and the calculated
light curves. The result is
130$\pm$30 km~s$^{-1}$~mag$^{-1}$, being somewhat larger 
then the mean value of 93 km~s$^{-1}$~mag$^{-1}$ found by Breger (1979).
This ratio depends on the non-adiabatic
behaviour of the modes and it is different depending on 
$n$ and $l$ values. That is why we do not find exact fit with the 
photometric solution. A much longer spectroscopic data series is
required to draw firm conclusions on the non-radial nature of oscillation.

By a close visual inspection of the individual spectra, we found the
metallic lines to show significant and highly variable asymmetries
(see Fig.\ 8).
The most straightforward explanation is the line profile distorsion
caused by non-radial pulsation. Line profile
variations among the multiply periodic $\delta$ Scuti stars
are found frequently and their analysis is a common method
of mode identification (see, e.g., Schrijvers et al. 1997,
Telting \& Schrijvers 1997 for theory and Mantegazza et al. 2000
for a recent application). Unfortunately, neither the resolution, nor
the relatively high noise level allow us to use such spectroscopic
methods, but the brightness of V784~Cas makes the star a good
target object for further spectroscopic investigations. 

As can be seen in Fig.\ 8, the spectral lines of V784~Cas are 
significantly broader than those of HD~187691 (for which
SIMBAD lists $v\sin i=3$ km~s$^{-1}$). Following the
work of Solano \& Fernley (1997), we have estimated the rotational
velocity of the star from the blend-free line Fe I 6677.997 \AA.
The resulting $v\sin i=55\pm10$ km~s$^{-1}$ is in agreement 
with the published value 66$\pm$10 km~s$^{-1}$ of De Medeiros \& Mayor (1999).
The most important point is that rapid rotation affecting 
the photometric parameters  (P\'erez Hern\'andez et al. 1999)
can be excluded.

Finally, we have to discuss the possible binarity, which may also produce
asymmetric line profiles when the partially resolved components are of similar
brightness. In our case the radial velocity measurements in the literature show
quite high scatter that may be associated with long-term orbital motion in an
unresolved binary system. We have searched the SIMBAD database for published
radial velocities and found the following data: {\it i)} Evans (1967) lists
$v_{\rm rad}$=+15 km~s$^{-1}$; {\it ii)} the catalog of Fehrenbach et al.
(1996) gives $v_{\rm rad}$=+20 km~s$^{-1}$; {\it iii)} the latter value has
been adapted in Duflot et al. (1995); {\it iv)} the first high-precision data
were published by De Medeiros \& Mayor (1999) giving $v_{\rm rad}=-$6.09
km~s$^{-1}$ (and it was used in L\`ebre \& De Medeiros 1997). Our mean value is
$\langle v_{\rm rad} \rangle =-6.3$ km~s$^{-1}$ supporting the latest available
data, which were taken in the early 90's. Although we do not know the
uncertainty of the early data, the $\approx$25 km~s$^{-1}$ difference found
seems to be too high to just ignore it.  Therefore, subsequent spectroscopy is
highly desirable, either the mode identification, or the possible binarity is
concerned. We note, that in light of the results presented in the next section,
we favour the non-radial pulsation and associate the large velocity difference
to the effects of metallic line profile distorsions. Furthermore, the lack  of
any variable asymmetry in the H$\alpha$ profile makes unlikely the presence of
a secondary of similar brightness as the primary one producing strong metallic
lines and having practically no contribution to the H$\alpha$ line. 
We did not find any change in the
systematic velocity between 1999  and 2001, that is why we consider non-radial
pulsation to be more likely cause of these asymmetries 
instead of a peculiar companion.

\begin{figure*}
\begin{center}
\leavevmode
\psfig{figure=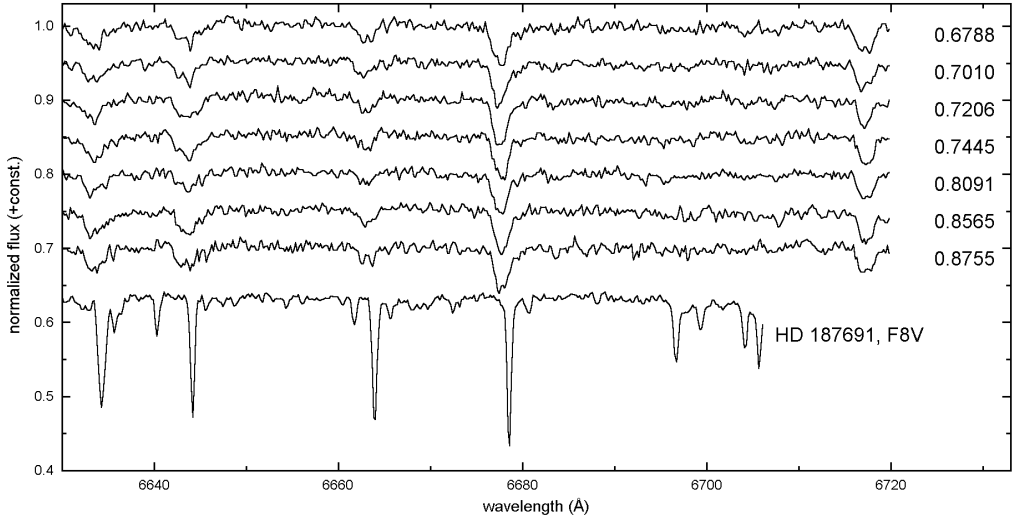,width=14cm}
\caption{Asymmetric metallic line profiles of V784~Cas. The bottom
spectrum is shown for comparison. The labels of the variable star
spectra mean $Hel.JD_{\rm obs}-2451480$.}
\end{center}
\label{f8}
\end{figure*}

\section{Physical parameters}

The most important physical parameters of V784~Cas were determined using the
mean Str\"omgren colours and the accurate Hipparcos parallax. Unfortunately,
there is no $\beta$ measurement for this star in the literature, therefore,
we could not follow the ``standard'' procedures of Str\"omgren photometric
calibrations (e.g. Zhou et al. 2001ab). The parallax is 9.81$\pm$0.75 mas
(ESA 1997) that corresponds to a distance of 102$^{+8}_{-7}$ pc.
Consequently, the reddening can be neglected even in this low galactic
latitude region ($b=-1.39^\circ$). The distance and mean $V$-brightness
result in $M_{\rm V}$=1\fm61$\pm$0\fm15. The spectral type (or,
equivalently, the mean $B-V$=0\fm33) implies a bolometric correction of
$BC=-0\fm11$ (Carroll \& Ostlie 1996). Thus, the
bolometric absolute magnitude is $M_{\rm bol}=1\fm50\pm$0\fm15
($L/L_\odot \approx$20), which is in accordance with the
expected absolute magnitude of an evolved early F-type star.
The atmospheric parameters were estimated from the mean
Johnson and Str\"omgren colours ($\langle b-y \rangle=0\fm20$,
$\langle m_1 \rangle=0\fm17$, $\langle c_1 \rangle=0\fm74$)
and synthetic colour grids of Kurucz (1993).
(We note, that the observed $uvby$ colours
are in very good agreement with those of listed in the
SIMBAD database: $\langle b-y \rangle=0\fm208$,
$\langle m_1 \rangle=0\fm204$, $\langle c_1 \rangle=0\fm735$ -- Olsen 1983).
The results are: $\langle T_{\rm eff} \rangle$=7100$\pm$100 K
and ${\rm log}~g$=3.8$\pm$0.1. For the given luminosity
and corresponding solar values ($M_{\rm bol, \odot}$=4\fm75,
$T_{\rm eff, \odot}$=5770 K, Allen 1976), the calculated stellar radius
is 2.9$\pm$0.3 R$_\odot$, while the estimated mass is 2.0$\pm$0.8 M$_\odot$.
Again, we find such stellar parameters that are typical for 
evolved main-sequence or giant stars (see, e.g., Appendix E
in Carroll \& Ostlie 1996). In summary, therefore, we adopt

\bigskip

\noindent $M_{\rm V}=1\fm61\pm0\fm15$

\noindent $M_{\rm bol}=1\fm50\pm0\fm15$

\noindent $L/L_\odot=20\pm3$

\noindent $\langle T_{\rm eff} \rangle = 7100\pm100$ K

\noindent $\langle {\rm log}~g \rangle =3.8\pm0.1$ dex

\noindent $R/R_\odot=2.9\pm0.3$

\noindent $M/M_\odot=2.0\pm0.8$

\bigskip

We can draw some constraints on the stellar mass and age using the
evolutionary tracks from Claret (1995) for solar abundances. They are
plotted in Fig.\ 9 with the corresponding position of V784~Cas in the
${\rm log}~T_{\rm eff}-{\rm log}~L$ plane. Two models with ${\rm
log}~M[M_\odot]=0.25-0.30$ are closest to the 
stellar error box and they result in an evolutionary mass of $1.89\pm0.11
M_\odot$. The corresponding ages range from 1.03 to 1.65 Gyr. 

\begin{figure}
\begin{center}
\leavevmode
\psfig{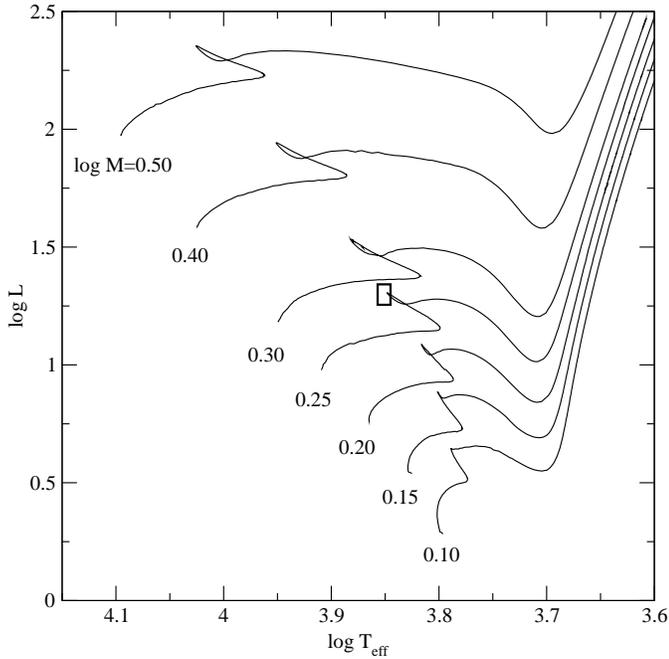}
\caption{Evolutionary tracks (Claret 1995) and location of V784~Cas 
in the HR diagram (thick box).}
\end{center}
\label{f9}
\end{figure}

The given parameters permit calculation of the pulsational constants
of the determined frequencies. The equation

$${\rm log}~Q=-6.456+{\rm log}~P+0.5{\rm log}~g+0.1 M_{\rm bol}+{\rm log}~T_{\rm eff}$$

\noindent was used in terms of four observables (Breger et al. 1993). For the
adopted set of four frequencies we have calculated the pulsational constants
listed in Table\ 4 (assuming 20\% uncertainty). We have also determined amplitude
and phase relations for the given frequencies utilizing Str\"omgren data obtained
in 2001. By fixing the frequencies but allowing their amplitudes and phases 
to vary for different wavebands, we could draw some contraints on the 
pulsation modes (see a description of the discrimination procedure in 
Garrido et al. 1990). For the given parameters, it seems to be clearly
established the radial character of $f_1$ and the non-radial character of 
$f_5$ (possibly $l$=2). We can say nothing definitive for $f_2$ and $f_3$
because errors are higher than values, but the small values seem to indicate $l$=1
modes. We note, that when the rotation is high enough the mode identification
becomes unclear (see Daszy\'nska-Daszkiewicz et al. 2002).

\begin{table*}
\caption{Pulsational constants and amplitude and phase relations
for the adopted set of frequencies. Note, that ordinal numbers are the same
as in Table\ 2.}
\begin{center}
\begin{tabular} {lrrrcccc}
\hline
i & $f_{\rm i}$ (c/d) & $P_{\rm i}$ (d) & $Q_{\rm i}$ & $v/y$ & $\phi_v-\phi_y$ & $b/y$ & $\phi_b-\phi_y$\\ 
\hline
1 & 9.1565  & 0.10921 & 0.030$\pm$0.006 & 1.61$\pm$0.03 & +5\fdg6$\pm$0\fdg5 & 1.29$\pm$0.02 & +3\fdg2$\pm$0\fdg5\\
2 & 9.4649  & 0.10565 & 0.029$\pm$0.006 & 1.58$\pm$0.07 & $-$1\fdg1$\pm$1\fdg4 & 1.27$\pm$0.06 & $-$1\fdg0$\pm$1\fdg4\\
3 & 15.4036 & 0.06492 & 0.018$\pm$0.003 & 1.46$\pm$0.09 & +1\fdg5$\pm$2\fdg0 &  1.20$\pm$0.08 & +1\fdg2$\pm$2\fdg0\\
5 & 15.9013 & 0.06289 & 0.017$\pm$0.003 & 1.55$\pm$0.13 & $-$8\fdg2$\pm$3\fdg1 & 1.25$\pm$0.13 & $-$5\fdg0$\pm$3\fdg1\\
\hline
\end{tabular}
\end{center}
\end{table*}

The parameters and pulsation pattern outlined above suggest V784~Cas to be
an evolved $\delta$ Scuti-type variable star with a 
mixture of radial plus non-radial modes. The star is presently about 1.4 mag
brighter than main sequence stars of the same spectral type (e.g. Carroll \&
Ostlie 1996) and its position above the main-sequence does not 
contradict the luminosity class III determined by Gray et al. (2001).
The star is located on the HR diagram about halfway between the theoretical
Blue Edge for radial overtones and the empirical Red Edge 
(see Fig.\ 1 in Breger \& Pamyatnykh 1998). Furthermore, comparing V784~Cas
with the evolved A$m$ stars in
Fig.\ 5 of Rodr\'\i guez \& Breger (2001), its position is also in 
agreement with the weak A$m$ nature suggested spectroscopically by Gray et al.(2001).
The temperature, surface gravity, mass and luminosity give a consistent picture
compared with the standard evolutionary models used by Breger \& Pamyatnykh
(1998). The determined physical parameters place V784~Cas in that region where
no fast evolutionary period changes are expected. Therefore, the slight phase shift of
$\approx-0.1$ between the Hipparcos data and our observations
(see Fig.\ 1) may indicate either more and yet undetected
pulsational frequencies or non-evolutionary period change
due to, for instance, light-time effect in a binary system
(Kiss \& Szatm\'ary 1995).

\section{Conclusions}

In this paper, we presented an analysis of photometric and spectroscopic
observations of the recently discovered $\delta$ Scuti variable V784~Cas.
The $UBV$ photometry was carried out at Szeged Observatory (Hungary), while 
the simultaneous $uvby$ data were obtained at Sierra Nevada Observatory
(Spain) in three consecutive years (1999-2001). Medium-resolution
spectroscopy in the H$\alpha$ region was carried out at the David Dunlap
Observatory (Canada) in 1999 and 2001. These data were used to determine
the frequency content and to estimate physical parameters of the star. The
main results can be summarized as follows:

\noindent 1. Multicolor data consisting of more than 3000 individual points
were analyzed with the standard Fourier-analysis. The multiperiodic nature
of the star is revealed unambiguously. Besides the dominant period listed
also in the Hipparcos catalog, we could detect three more frequencies in the
9.46-15.9 d$^{-1}$ range. There is a suggestion for more, unresolved frequency
components.

\noindent 2. We have obtained almost 100 radial velocity measurements using
the H$\alpha$ line. The measured radial velocity curves also show the
multiperiodic nature and a close correlation with the four-component
light curve fit. Spectra obtained in 1999 covered a few weak metallic lines
and the varying asymmetric line profiles suggest the presence of 
non-radial pulsation, too.

\noindent 3. Physical parameters of the star are determined from the mean
Str\"omgren indices and synthetic colour grids. The resulting parameters
give a consistent picture of an evolved $\delta$ Scuti star.
Evolutionary mass ($1.89\pm0.11 M_\odot$) and age ($1.3\pm0.3$ Gyr) is derived.

\noindent 4. Possible mode identification was discussed 
based on the Str\"omgren photometric behaviour (amplitude and phase
relations). We identify $f_1$ with the radial fundamental mode,
while the remaining frequencies correspond to low-order ($l$=1 or 2)
non-radial modes, although some ambiguity may arise from the 
moderate rotation of the star.

Further observations (photometric, as well as spectroscopic) of this
variable star are expected to extend the data baseline yielding to a 
better resolution of the pulsational pattern, mode identification 
and detection of time-dependent phenomena (e.g. amplitude and/or frequency
modulation).

\begin{acknowledgements}
This work has been supported by the MTA-CSIC Joint Project No. 15/1998, OTKA
Grants \#T032258, \#T030743 and \#T034615, the ``Bolyai J\'anos'' Research 
Scholarship to LLK from the Hungarian Academy of Sciences, the
Hungarian E\"otv\"os Fellowship to LLK, FKFP Grant 0010/2001,
Pro Renovanda Cultura Hungariae Grant DCS 2001/\'apr/6. and Szeged Observatory
Foundation. Fruitful discussions with J. Vink\'o are
acknowledged. The NASA ADS Abstract Service was used to access data and
references. This research has made use of the SIMBAD database, operated at
CDS-Strasbourg, France.
\end{acknowledgements}

\end{document}